\documentstyle[12pt,epsfig]{article}
\begin{document}
\thispagestyle{empty}

\hfill{\sf CERN-TH/2001-020}

%\hfill{\sf Presented at Hot Points in Astrophysics}

%\hfill{\sf JINR, Dubna, Russia, August 22-26, 2000}

\vspace*{3.cm}
\begin{center}
{\large \bf Big Bang Nucleosynthesis and Cosmological Constraints\\  
on Neutrino Oscillation Parameters}
\end{center}
\vspace*{0.2cm}
\begin{center}
Daniela Kirilova$^{a,c}$ and Mihail Chizhov$^{b,c}$\\[0.3cm]
{\it $^{a}$Institute of Astronomy,Bulgarian Academy of Sciences, 
Sofia, Bulgaria\\
$^{b}$Centre of Space Research and Technologies, 
Sofia University, Sofia, Bulgaria\\
%and 
$^{c}$Theory Division, CERN, Geneva, Switzerland}
\end{center}
\vspace*{0.2cm}
\begin{abstract}
  We present a review of cosmological nucleosynthesis (CN)
with neutrino oscillations, discussing the different 
effects of oscillations on CN, namely:  increase of the
effective degrees of freedom during CN, 
 spectrum distortion of the oscillating neutrinos,
 neutrino number density depletion,  and
  growth of neutrino--antineutrino asymmetry due to
active--sterile oscillations. 
We discuss the importance of these effects for the 
primordial yield of helium-4. 

Primordially produced $^4$He value  
is obtained in a selfconsistent study of the nucleons and the
oscillating neutrinos. The effects of spectrum distortion, depletion and
neutrino--antineutrino  asymmetry growth on helium-4 production are
explicitly calculated.

An update of the cosmological
constraints on active--sterile neutrino oscillations  parameters is 
presented, giving the values: $\delta m^2 (\sin^22\vartheta)^4\le 1.5
\times10^{-9}$ eV$^2$ 
for $\delta m^2>0$, and $|\delta m^2| < 8.2\times 10^{-10}$ eV$^2$
at large mixing angles for $\delta m^2<0$.   
According to these constraints, besides the active--sterile 
LMA solution, also
the active--sterile LOW solution to the solar
neutrino problem is almost totally excluded.
\end{abstract}

\vspace*{0.5cm}

\newpage

\section{Introduction}
  Cosmological nucleosynthesis is traditionally used as a probe of the
conditions of the early Universe at the nucleosynthesis epoch 
(T $\sim$ MeV). Concerning neutrino physics, the requirement for a
concordance
between the theoretically predicted and the extracted from observations
primordial abundances of light elements, constrains 
neutrino characteristics: mass, number of light  
species, degeneracy,  decay width and mass of eventual heavy 
decaying neutrinos, neutrino oscillation parameters, possible new
interactions, etc.  

On the other hand, the topic of  neutrino oscillations is with us since
Pontecorvo's hypothesis for these~\cite{pontec} i.e. more than 40 years. 
Neutrino oscillations imply non-zero neutrino masses and mixings,
therefore presenting  an indication of physics beyond the
electroweak standard model. Recently 
the positive indications of oscillations obtained 
from the greatest neutrino
experiments (SuperKamiokande, SNO, Soudan 2, LSND, etc.)~\footnote{
  All recent data of {\it the solar neutrino} experiments
namely Gallium, Chlorine, SuperKamiokande, SNO, sensitive to
different
neutrino energy,  point to a well pronounced
deficit of solar neutrinos in contrast with the theoretical
predictions of the Standard Solar Model fluxes.

  The measured ratio of the muon over electron neutrino flux by
Kamiokande, SuperKamiokande, IMB and Soudan 2 is considerably lower than
the predicted one for {\it the atmospheric neutrinos}, resultant from the 
primary cosmic rays collisions with the nuclei in the upper atmosphere.
Moreover, a zenith-angle-dependent deficit of muon neutrinos was observed. 

  {\it The LSND experiment} has observed electron antineutrino appearance in
a flux of muon  antineutrinos and electron neutrinos in a flux of
muon neutrinos. 

These three neutrino anomalies are preferably explained  in
terms of neutrino oscillations, which are able to provide a natural
mechanism for energy dependent and neutrino type dependent
suppression of neutrino fluxes.
There exist several solutions to the solar neutrino problem:
Small Mixing Angle (SMA), Large Mixing Angle (LMA), LOW and 
Vacuum Oscillations (VO) solutions,  
which imply neutrino mass differences in the range: $10^{-10} - 10^{-4}$
eV$^2$. The atmospheric neutrino anomaly can be resolved by 
oscillations 
with mass differences $10^{-3} - 10^{-2}$ eV$^2$. While for the LSND
neutrino experiment, much bigger mass difference, of the order of eV is
necessary.}
turned the subject of neutrino oscillations
into one of the hottest points of astrophysics and neutrino physics.   
Hence, it looks like appropriate to provide 
 an updated review of the influence of    
 neutrino oscillations  on CN and present the most recent cosmological 
 constraints on neutrino oscillations parameters.

In case neutrino oscillations are present in  the primordial plasma
of the early Universe, they may lead to changes in CN, depending on the
 oscillation channels and the way they proceed. Namely, the 
oscillations effect depends on the kind of oscillations (they can
be resonant or nonresonant) and also differs for the    
equilibrium and  nonequilibrium  cases.
Oscillations may influence  nucleosynthesis
through their effects on:
 neutrino and antineutrino number densities,
 spectrum, neutrino-antineutrino asymmetry and   
 the number of neutrino species. 

In the next section we will provide a review of CN
with neutrino oscillations, discussing mainly oscillations 
influence  on CN. 
In the last section we will present an update of the cosmological
constraints on $\nu_e \leftrightarrow \nu_s$ neutrino oscillation
parameters and discuss how they concern the solutions to the 
solar neutrino problem.

\section{Historical overview of CN with neutrino oscillations}

\subsection{Standard Big Bang Nucleosynthesis}

  For a  precise analysis of
the oscillations effect on CN, 
the element that is used traditionally
is helium-4, as far as the most
reliable and abundant data now available are for that element.
According to the standard Big Bang Nucleosynthesis (SBBN) 
theory $^4$He is
a result of a complex network of nuclear reactions, which
proceed after the  freezing of the neutron-to-proton ratio $n/p$.

  The abundance
of the primordially produced mass fraction of helium-4 
$Y_p\sim2 (n/p)_f/$ $(1+(n/p)_f)$ depends 
mostly on
two compelling   
processes, determining the neutron-to-proton freezing ratio -- $(n/p)_f$,
namely the Universe's
cooling rate, $H(t)\sim \sqrt{g_{eff}}~T^2$ and the
interaction rates of the weak processes, interchanging neutrons and protons: 
 $\Gamma_w \sim G_F^2(g_V^2+3g_A^2) E_{\nu}^2 T^3$.
 Hence, the produced helium is a strong function of the
 effective number of relativistic degrees of freedom at the CN epoch,
$g_{eff}$, and the neutron mean lifetime $\tau_{n}$, 
which parametrizes~\footnote{
Where $\tau_n^{-1}\sim G_F^2(g_V^2+3g_A^2)$.}
the weak interactions strength. Besides, primordially produced 
$^4$He is a
logarithmic function of the baryon-to-photon ratio $\eta$, due to the  
nuclear reactions dependence on  nucleon densities, i.e. 
$Y_p(g_{eff},\tau_{n},\eta)$. 
Deuterium  measurements in  pristine
environments towards  low metallicity quasar absorption systems  at very
high $z \sim 3$~\cite{maera}
 provide us with the most precision determination of the baryon density,
giving the value: $\eta=5.6 \pm 0.5 \times 10^{-10}$. Recently, 
the baryon density was also determined  
 from observations of the anisotropy of the cosmic
microwave background (CMB) 
 by DASI, BOOMERANG and  MAXIMA experiments~\cite{etaCMB}.
 The CMB value is  in
agreement with the one found from deuterium measurements and SBBN.

{\it SBBN assumes three neutrino flavours, zero lepton asymmetry and 
equilibrium neutrino number densities and spectrum:}
\begin{eqnarray*}
n_{\nu}^{eq}=\exp(-E/T)/[1+\exp(-E/T)]
\end{eqnarray*}
The primordial $^4$He  abundance $Y_p$, predicted from SBBN, is
calculated
with  great precision. In recent years, corrections to  SBBN
code accounting for  different physical effects reduced  the theoretical
uncertainty
to less than  0.1\% ($|\delta Y_p| < 0.0002$) within a wide range of
$\eta$~\cite{helium}. 
The uncertainty of the observational $Y_p$ is few percent.
The predicted helium-4 value 
is in accordance with the contemporary helium values, inferred from
observational data: 0.238--0.245 (the  systematic errors are supposed
to be around $0.007$)~\cite{heliumd}, and is consistent with other light
elements abundances.
%\footnote{ The predicted $Y_p$ at the  best fit
%value of $\eta$ obtained from deuterium measurements 
%$\eta=5\times 10^{-10}$, is $Y_p=0.2462$.}.

Given this accuracy, it  
can be used as a probe of the eventual new neutrino
physics -- the neutrino oscillations.

\subsection{CN with neutrino oscillations}  

CN with neutrino oscillations was studied in
numerous publications~\cite{do}--\cite{bern}. 
The basic idea of oscillations is 
that neutrinos are not degenerate in mass and there is a
mixing, so that the mass eigen-states $\nu_i$ are distinct from
the flavour eigen-states $\nu_f$ and are unitary combinations
of the latter: 
$$
\nu_i=U_{if}~\nu_f~~~~~ (f=e,\mu, \tau).
$$

Then in the simple two-neutrino oscillation case
\begin{eqnarray*}
\nu_1 & = & c\nu_f+s\nu_{f'},\\
\nu_2 & = & -s\nu_f+c\nu_{f'},
\end{eqnarray*}
where 
$c=\cos(\vartheta)$,
$s=\sin(\vartheta)$, $\vartheta\neq 0$ is the mixing angle, 
$\nu_1$ and $\nu_2$ are
the mass eigen-states  with masses correspondingly $m_1$ and
$m_2$ and $\delta m^2 \neq 0$.

 The probability to find 
at a distance $l$
a given neutrino type $f'$ in an initially homogeneous neutrino beam of
another neutrino type $f$ is:
$$
P_{ff'}=\sin^22\vartheta \sin^2 (1.267~\delta m^2 l/E),
$$
\noindent where $\delta m^2$ is the neutrino mass difference in eV$^2$, 
 $E$ is the neutrino energy in MeV and $l$ is the distance in meters. 
 
\subsubsection{Vacuum oscillations}

The oscillations effect on CN was historically first considered
for vacuum {\it oscillations between 
different neutrino flavours}~\cite{do,hp}. 
It was shown that since there is a slight deviation
from equilibrium in that case (temperatures of different flavour neutrinos
are nearly the same), flavour oscillations have a negligible effect on
neutrino number densities and on nucleosynthesis~\cite{do}.
The possibility of generation of neutrino-antineutrino asymmetry in 
the electron sector due to CP-violating oscillations and its effect on 
CN was discussed in ref.~\cite{hp}.

%{\it Active--sterile neutrino~\footnote{$SU_2\times U_1$ singlet neutrino}
%!!!
{\it Active--sterile neutrino~\footnote{$SU_2$ singlet neutrino.}
oscillations} 
effect on CN may be noticeably stronger
because they can 

(a) increase the effective number of light degrees of 
freedom during nucleosynthesis~\cite{do,ns,la89};    

(b) lead to a strong distortion of the neutrino spectrum~\cite{dpk}; 

(c) lead to a considerable decrease of the total number density
of active neutrinos~\cite{dpk}; 

Active-sterile oscillations may keep sterile neutrinos in thermal
equilibrium  or bring them into equilibrium in case they have 
already decoupled. CN allows not more than one additional 
neutrino type, therefore forbidding efficient production 
of sterile neutrinos due to oscillations ($\Gamma_R<H$). 
First constraints~\cite{do} on oscillation parameters were obtained  
from the requirement that the observed primordial values   
allow less than one additional light neutrino type.
It was shown that the simultaneous presence of Dirac and Majorana 
neutrino masses would contradict the observed helium-4 abundance 
if $\delta m^2 > 10^{-6}$ eV$^2$. 

In ref.~\cite{ns} CN constraints were estimated 
for fast oscillations ($\Gamma_{osc}>H$) from the condition that 
sterile neutrinos should not be abundantly produced due to oscillations 
before nucleosynthesis epoch  ($\Gamma_R\sim \Gamma_w
\sin^22\vartheta <H$ at 1 MeV):
$(\delta m^2)^2 \sin^22\vartheta\le1.6\times 10^{-17} eV^4$. 

In ref.~\cite{la89} a condition for an efficient production of sterile
neutrinos
% $\Gamma_wP_{es} > H$
 at  about 1 MeV,
accounting for the rescattering of neutrinos from the
fermions of the plasma, was calculated and constraints on neutrino
oscillation parameters were also obtained. These constraints 
~\cite{ns,la89} as will be 
discussed in more detail in the next subsection, are considerably relaxed, 
when a proper account of the dispersion effects is provided. 

In ref.~\cite{dpk} the nonequilibrium case of 
active--sterile oscillations,  effective after the
neutrino decoupling, i.e.  
  the oscillation rate
exceeding  the expansion rate $\Gamma_{osc}\sim\delta m^2/(4E)>H$, 
while neutrino weak rates less than expansion rate 
 $\Gamma_w<H$, was discussed.

The effect of such active-sterile neutrino oscillations  
is due mainly to the fact that sterile neutrinos
may have decoupled
much earlier than the active ones~\footnote{Therefore, they were not
heated by the following after their decoupling processes leading to
entropy increase.} and
at the nucleosynthesis epoch
their number densities were  negligible in comparison with the
active neutrinos: $N_s \ll N_f$, as far as $T_s<T_f$ and $N\sim T^3$.

The accurate kinetic approach of ref.~\cite{do} to a
description of the oscillating 
neutrinos in terms of {\it neutrino  density matrix in momentum space},
was used in ref.~\cite{dpk} to calculate analytically 
the evolution of the number density of electron neutrinos:
$$
n_{\nu_e}=\rho_{LL}=
\left\{1-2c^2s^2+2c^2s^2\cos [BT/E(T^{-3}-T_0^{-3})]\right\}\times
n_{\nu_e}^{eq},
$$
\noindent where $B=0.1 M_{Pl}\delta m^2/\sqrt{g_{eff}}$ and $T_0=3$ MeV.
 
It was explicitly shown,  that
for large interval of oscillation parameters of the 
discussed model, oscillations
may cause considerable spectrum distortion and/or   
depletion of the electron neutrino. 

At great $\delta m^2$,  $\cos [BT/E(T^{-3}-T_0^{-3})]$
is frequently oscillating and can be averaged. Then  
$\rho_{LL}=(1-2c^2s^2)n_{\nu_e}^{eq}$, and   
active neutrinos number densities are  depleted by oscillations. 
However, for smaller $\delta m^2$, the energy distribution of the active
neutrinos 
may be also considerably changed.

Both the electron neutrino depletion and the decrease of the 
electron neutrino energy due to oscillations into less energetic sterile 
neutrinos decrease the weak rates $\Gamma_w \sim E_{\nu}^2 N_{\nu}$, 
leading to higher freezing temperature of the nucleons.
The total effect is an overproduction of He-4.

Helium production was numerically 
calculated for 
the full range of parameters of this oscillation model.
It was shown that,
the  effect of depletion and spectrum distortion due to oscillations 
 on He-4  may be  much stronger  than the  
effect due to excitation of an additional degree of freedom. 
 Assuming that allowed overproduction should  be less than $5\%$, 
the following constraints on the oscillation parameters were obtained:
          $\delta m^2<10^{-9}$ at $\vartheta>\pi/15$ 
The overproduction of helium due to sterile  
neutrino species excited by oscillations was
calculated. 

\subsubsection{Matter oscillations}

In the following years, it was realized that for a large interval of
masses and mixing angles the thermal background in 
the pre-nucleosynthesis epoch may strongly affect the propagation of
neutrinos  and, has to be taken into account.
Differences in the interactions
with the particles from the plasma lead to different average potentials
for different neutrino types~\cite{nr}. 
In the adiabatic case
the effect of the medium can be formally hidden in the oscillation
parameters, by introducing matter oscillation parameters that 
are expressed through the vacuum ones and through the characteristics of
the medium. 
The matter mixing angle is then~\cite{nr,MSW}
$$
\sin^2\vartheta_m=\sin^2\vartheta/[\sin^2\vartheta+(Q \mp
L-\cos2\vartheta)^2],
$$
\noindent where $Q=-bE^2T^4/(\delta m^2M_W^2)$, 
$L=-aET^3L^{\alpha}/(\delta m^2)$,
$L^{\alpha}$ is expressed through the fermion asymmetries of the plasma, 
$a$ and $b$ are positive
constants different for the different neutrino types,
$-L$ corresponds to the neutrino and $+L$ to 
the antineutrino case.

It was realized that although most often the medium suppresses
oscillations (by  decreasing their amplitude),
there also exists  a possibility of enhanced 
oscillation transfer in case a resonant condition between the 
parameters of the medium
and the oscillation parameters holds: 
$$
Q \mp L=\cos2\vartheta.
$$
\noindent Then the mixing in matter becomes maximal, independently of the
value of the vacuum mixing angle. 
Besides, as can be easily judged from the resonant condition, 
at high temperatures, when $|Q|>|L|$, $\delta m^2>0$ corresponds
to a nonresonant case, 
while $\delta m^2<0$ corresponds to a resonant case, and the 
resonance 
holds in both  neutrino and antineutrino sectors. 
At low temperatures, when $|Q|<|L|$, the resonance is possible 
either for neutrinos
in the case $\delta m^2>0$ or for antineutrinos in the 
case $\delta m^2<0$.

{\it Matter oscillations of ordinary flavour neutrinos}, occurring before 
the freeze-out of $n/p$ - ratio were considered first~\cite{la}. 
Two different effects of oscillations were taken into account: 
the generation of $\nu_e-\bar{\nu_e}$ asymmetry in  
$\nu_{\tau}\leftrightarrow \nu_e$ oscillations resonant conversion 
and the change in the neutrino freeze-out temperature. 
The effect on primordial production of He-4 was estimated to 
be very low: $\delta Y_p<1.3\times 10^{-3}$.

The effect of {\it matter active--sterile oscillations} 
   on CN, taking into account the essential processes of 
neutrino forward scattering of the background particles, 
was  considered first in
refs.~\cite{smirnov},\cite{bd1}-\cite{bd2},\cite{ekt}.
Active--sterile oscillations that are efficient before 
the nucleosynthesis epoch (corresponding to high mass differences) 
were discussed there.
The production of sterile neutrino states due to oscillations and
interactions of the active neutrino with the medium was calculated. 

It was shown that matter oscillations, 
proceeding before the active neutrinos decoupling, are capable of   
 exciting additional degrees of freedom into the plasma. The  
increase in the
expansion rate $H \sim \sqrt{g_{eff}}$ and the corresponding
overproduction of helium were calculated. 

The exclusion regions for the 
neutrino-mixing parameters were obtained from the requirement that the 
number of neutrino species is less than a certain value 
(usually $N_{\nu}<3.4$ in the pioneer publications), up to which there is 
an agreement between the values for light element abundances 
obtained from theoretical predictions  and those extracted from
observations.  The excluded region of 
oscillation parameters values was found  much smaller than in the case 
not accounting for the neutrino coherent interactions.

 In these works  the kinetic evolution of the 
neutrino ensembles was studied in terms of particle densities, and  it
was assumed that oscillations
slightly shift  neutrino density matrix from its diagonal form. 
First estimates of an eventual depletion of electron neutrinos, 
due to  $\nu_e \leftrightarrow \nu_s$ oscillations, and its corresponding
influence on helium-4 were made in~\cite{bd1,bd2}.
 Numerical calculations of the 
discussed oscillations effects on helium-4 production were 
made in~\cite{ekt}. 

The idea that active-sterile oscillations will produce considerable 
$\nu_e-\bar{\nu_e}$ asymmetry during resonant transfer, which may
influence CN was discussed in ref.~\cite{smirnov}.

\subsubsection{CN constraints on active-sterile oscillations}

The pioneer works discussing $\nu_e\leftrightarrow \nu_s$
 at  $\delta m^2>0$ excluded the Large Mixing
Angle solution (LMA) of the solar neutrino problem.

In the $\nu_{\mu}\leftrightarrow \nu_s$ oscillations case, 
the solution to the atmospheric 
anomaly was found to lie in the cosmologically excluded region.
This can be regarded as an indication that the atmospheric neutrino
anomaly should be solved by oscillations in the $\nu_{\mu}
\leftrightarrow \nu_{\tau}$ sector 
(which is also the experimentally preferred recent 
solution)~\footnote{With 
this in mind, and also the fact that the asymmetry
growth due to resonant oscillations still
needs some clarification and precision calculations, we will not 
 discuss the numerous
works aiming to save the active--sterile solution, by
weakening
the cosmological constraints due to the large asymmetry production in
active--sterile oscillations in the muon or the tau-neutrino sector.}.

Recently~\cite{do00} the constraints on active--sterile neutrino
oscillations
were discussed in connection with the controversy in literature
concerning the sterile neutrino production rate: in some works
annihilation rate was used when calculating it, while in others the
total reaction rate was employed. The problem is still under discussion.

In ref.~\cite{ssf} (see also ref.~\cite{shif99})  a  rough
estimation of the spectrum distortion
by shifting the effective temperature of the neutrino and considering its
spectrum equilibrium was provided. 
 However, as was shown in refs.~\cite{NU96,PR,NP,res}   
this way  the real 
distortion of the neutrino spectrum cannot be described. 
Besides, the exclusion plots  of refs.~\cite{ssf,shif99} for 
the resonant case are not compatible with the ones for  the nonresonant
case, they 
do not coincide at
maximal mixing, as they should. Hence, it is preferable to 
 use the results of ref.~\cite{ekt} for the cosmological
constraints in the $\nu_{\mu}
\leftrightarrow \nu_s$ and $\nu_{\tau}
\leftrightarrow \nu_s$ as more reliable.

 It should be kept in mind also that  the constraints on
$\nu_{\mu,\tau}\leftrightarrow\nu_s$ have not  been
updated:  they do not take into account 
the spectrum distortion of the oscillating neutrinos and  the growth 
of the neutrino--antineutrino asymmetry. 

In the resonant case we expect the bounds to  become less restrictive
in  the small mixing angles area, 
when the estimation of the asymmetry growth due to oscillations is
made, because oscillations generated asymmetry at small mixing angles 
leads to a suppression of oscillations and consequently to 
a relaxation of the bound on oscillation parameters at small 
mixings~\cite{CAPP}. In both
resonant and nonresonant cases we expect less restrictive 
bounds at large mixings due to spectrum distortion effect.

%The effect of the spectrum distortion is a synthetic one: 
%it will influence both the  expansion
%rate and  the neutrino weak-interaction rates. 
%the resultant effect on He-4 should be precisely calculated.

\subsection{Update of oscillations effects on CN}

In the last  years the oscillations effects on CN  
were updated ~\cite{NU96}--\cite{res},\cite{now2000}, 
by providing a  precise
numerical account for the neutrino 
spectrum distortion, depletion  and  growth of neutrino-antineutrino 
asymmetry due to 
neutrino oscillations.

\subsubsection{Spectral distortion.}

In refs.~\cite{NU96}--\cite{res},\cite{now2000} exact kinetic
equations
 for the
neutrino density matrix in
{\it momentum} space, 
accounting {\it simultaneously} for Universe expansion,
neutrino oscillations and neutrino forward scattering, were 
numerically solved for oscillations with small 
mass differences. The equations 
 for the
neutrino density matrix  were solved {\it selfconsistently}  with the
kinetic 
equations of the nucleons during the $n/p$ - freezing (see detail 
description in the next section), which  
 allowed to provide a precise analysis of neutrino depletion 
and spectrum distortion effect on CN due to  oscillations. 
The spectrum distortion effect was shown to be considerable~\cite{NU96}.
In Fig.~1 the net effect of spectrum distortion is given by
the difference between the short-dashed curve from ref.~\cite{ekt} and
the long-dashed curve from ref.~\cite{res}. 
In the nonequilibrium case of electron-sterile oscillations, it  cannot
be  estimated simply by shifting the effective temperature or 
introducing chemical potentials. 

\begin{figure}[h]
\begin{center}
\epsfig{file=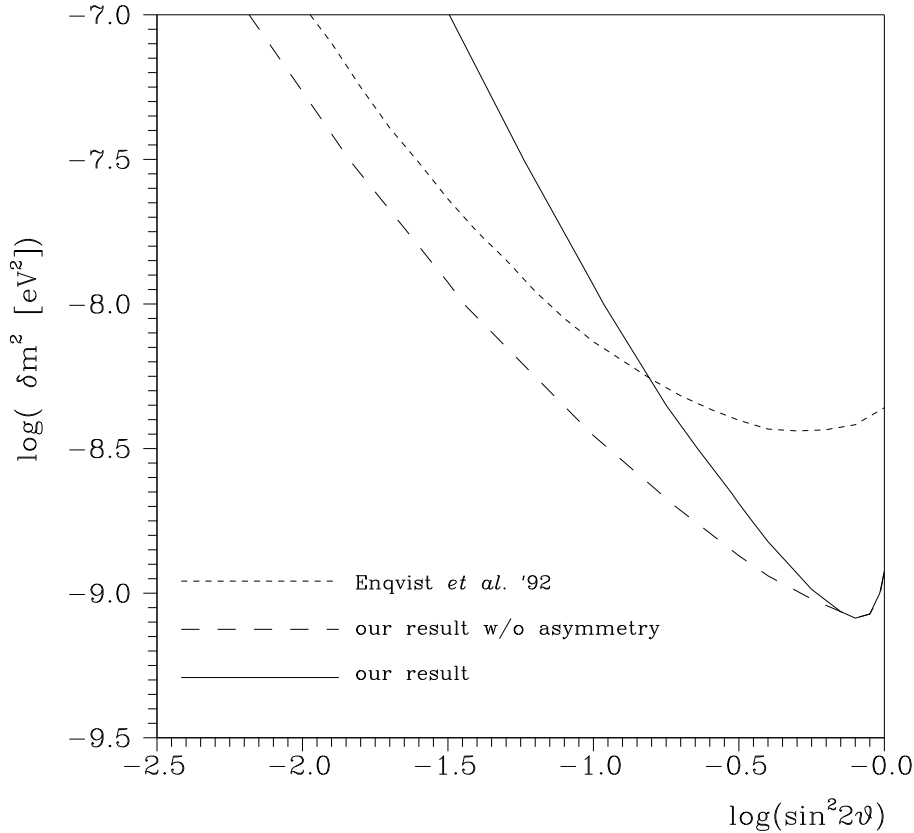,height=11cm,width=12cm}
\end{center}
\vspace{-0.3cm}
\caption{\small Comparison between the results on
primordial helium-4 production obtained in our works~\cite{new,res}
and in previous work~\cite{ekt}.
The long-dashed curve shows our results when
the asymmetry effect is neglected. The short-dashed curve
shows the results of Enqvist et al., where both asymmetry
and spectrum distortion effects were neglected.
The difference between the two curves shows  the effect
of the proper account of the spectrum distortion of the neutrino.
On the other hand, the difference between our curves, the solid and
the dashed one, presents the net asymmetry effect.}
\vspace{-0.3cm}   
\end{figure}

On the other hand the precise numerical analysis of the spectrum
distortion 
is technically very difficult.
It has been  provided only for  oscillations with
small mass
differences~\cite{NU96,PR,res}. For the adequate description of the
spectrum distortion in
that case  
1000 bins  in
the nonresonant case  and up to 5000 bins for the resonant case were
needed. Hence, thousands times larger system of coupled
integro-differential
equations had to be accurately solved, in contrast to the case of momentum 
averaged calculations when a set of 8 ordinary differential equations  
are solved. The  
computational time was strongly increased.  

Recently a successful analytical description of the spectrum distortion 
of neutrino spectrum due to oscillations was proposed in
ref.~\cite{asym} for small 
mixing angles and large mass differences. It may be very useful in
precision calculations of  the oscillations effect on CN  
at large mass differences, which have not been done.  

\subsubsection{Neutrino-antineutrino asymmetry growth}
  
The idea of generating neutrino-antineutrino asymmetry 
due to resonant transfer of neutrinos in active-sterile neutrino oscillations 
was first 
proposed in ref.~\cite{smirnov}. 
Alas, it was not developed further by the authors and remained unnoticed. 

Estimations of the effect of 
{\it asymmetry, generated in flavor oscillations}, on CN of helium-4 were 
provided in ref.~\cite{la}. The effect was shown to be negligible.   

Oscillations generated asymmetry  was  discussed in  more detail for {\it 
active-sterile  
oscillations} with large mass differences in 
refs.~\cite{bd1}-\cite{bd2}, where it was estimated to have 
negligible effect on CN, and in ref.~\cite{create}, where 
 asymmetry growth was found possible in case 
$\delta m^2>10^{-5}$ eV$^2$. In the  numerical
study,  exploring oscillations with
small mass differences  $\delta m^2<10^{-7}$ eV$^2$~\cite{NU96}, a growth
of asymmetry was registered. 

The revealment of the asymmetry
growth at small
mass differences became possible thanks to the exact
kinetic approach, namely work with selfconsistent kinetic
equations for neutrinos in momentum space. This approach  allows
to take into account  the
neutrino spectrum distortion and describe the
asymmetry behaviour  at each momentum.

In ref.~\cite{new} the evolution of the asymmetry was studied
selfconsistently with the neutrino
and nucleon evolution, which allowed to take into account the
effect of
the asymmetry on CN during {\it the full evolution of the
asymmetry}. 
It was shown that besides the well-known kinetic effect of the asymmetry
on CN, 
after it has grown up to high values of the order of $\pm 0.01$, 
very small asymmetries $L< 0.01$ exert a considerable {\it indirect
effect on CN through oscillations}. 
This is due to the fact that even very small asymmetries change the 
medium-induced neutrino potential energy and may have a considerable back 
reaction on  the
oscillating neutrinos. 
 It  has been shown~\cite{new,now2000}, that the results obtained when the
indirect asymmetry effect was taken into account differ by
 many orders of magnitude from results neglecting small asymmetries.

For the case of small mass differences exact kinetic calculations 
of the asymmetry effect 
were provided in refs.~\cite{NU96}--\cite{CAPP}. 
The asymmetry growth up to 5 orders of magnitude was registered
in these studies.
Hence,  these
lepton asymmetries experienced {\it indirect} effect on CN through changing
neutrino number densities and the oscillation pattern, which on its turn
effected the kinetics of nucleons during $n/p$ - freezing.

 The net {\it indirect} asymmetry
effect on CN is illustrated in Fig.~1  by the
difference between the solid and the long-dashed curves.
The asymmetry growth reduced neutrino 
 oscillations effect  and lead up to  $10\%$ underproduction of helium in
respect to the model of CN with oscillations but neglecting 
asymmetry. Hence dynamically produced asymmetry  leads to a noticeable
weakening 
 of the cosmological constraints for small mixing
angles~\cite{NU96,new,res}.

However, the precise numerical description of  the asymmetry
growth requires more than 5 000    
bins for the spectrum distortion, which was found  essential for 
the correct calculation of the asymmetry evolution~\cite{res}.  
This increases enormously the calculational time in the resonant case. 
Besides, the neutrino evolution
equations at resonance have  high stiffness and
implicit methods should be used
to solve the equations numerically.
For 5 000 bins of the spectrum  a system of 30 000
integro-differential equations describing the
neutrino density evolution should be solved simultaneously.

 Hopefully, the asymmetry effect on CN for large mass differences may be  
successfully analytically estimated using the                      
analytical
approximations, accounting  for the spectrum distortion
of neutrino, like the one proposed in
~\cite{asym} for small mixing angles and large mass differences.
Such approximations  
 maybe very useful for analyzing oscillations effect on CN, because they
simplify the equations  and reduce the computational time.

\subsubsection{Relic lepton asymmetries.}

The case of different {\it initial asymmetries}, namely $10^{-10}<L<10^{-4}$, 
in models of CN with {\it nonresonant} oscillations, effective after 
electron neutrino freeze-out, was
precisely studied
in~\cite{PR,NP,CAPP,now2000}. It was found that 
depending on the concrete values of the relic  asymmetry, it may
suppress, enhance or not influence
oscillations, and hence lead to an under-, overproduction of helium
or to not changing its abundance at all.

The possibility of lepton asymmetry to enhance oscillations,
besides
its well-known ability to suppress them, was 
revealed thanks to the study of neutrino evolution 
at each momentum: 

For the parameters of the  model 
  the local term $L$ dominates
over the other terms for the mean neutrino momenta, and hence the
resonant condition is not fulfilled:
$|L|\gg|Q|$, $|L|\gg\cos 2\vartheta$.

However, for  neutrinos with a given momentum $p<\bar{p}$ it 
is fulfilled and these neutrinos suffer a resonant transfer 
leading to the decrease of their number densities and a decrease 
of the asymmetry (since the resonance condition 
is fulfilled only in one sector, either neutrino or neutrino one, 
the particle densities in the other sector do not change).
Then, due to the  $L$ decrease, resonant transfer becomes possible  for
neutrinos with higher momentum, and so on till $L$ changes sign 
and the running resonance wave similarly passes through the
 antineutrino ensemble, leading to a rapid increase of $L$ again 
till the next change of sign of $L$,  etc. 
i.e. the observed enhancement is ``spectrum resonance'' effect.  

The total effect of the
process   is enhanced resonance transfer both in neutrino and
antineutrino sectors, leading to greater overproduction of helium-4. 
Oscillation constraints for the case of CN with
oscillations and with initial $L$ were obtained~\cite{NP}.

A similar investigation for the resonant case will be more complicated
 due to technical problems because it will deserve much greater number of
bins for the spectrum distortion description because of the asymmetry 
growth, and hence, longer calculational time.
However, this investigation is interesting, as far as such small values of
the initial relic asymmetry  are not excluded neither from observations   
nor from some profound theoretical principle.

In conclusion we would like to stress that even a very small asymmetry,
 either initially present or dynamically generated,
thanks to its indirect effect through
neutrino oscillations is capable to  influence strongly CN.
Hence it should be accounted for as precisely as possible. 
  
\subsection{Summary of neutrino oscillations effect on CN} 

In case neutrino oscillations between active and sterile neutrinos
proceed in the primordial plasma during
the CN epoch, they can effect CN in the following ways:

(a) In case active neutrinos have not decoupled, their oscillations to
sterile ones  may bring additional degrees of freedom
 into the primordial heat bath. This will lead to an increase of  the
Universe expansion rate H(t) and
to an earlier freezing of the $n/p$ - ratio, at times when
neutrons were more abundant in comparison with the SBBN.
 Hence, this effect leads to an overproduction of helium-4.
This effect was historically the first discussed
~\cite{ns} and on its basis first
constraints on the oscillation parameters $\delta m^2, \sin^2(2
\vartheta)$ were estimated~\cite{la89,bd1,ka,kimmo,bd2,ekt}.

(b) Oscillations considerably influence  CN by 
distorting  the neutrino and antineutrino spectrum. As far as the
oscillation rate is energy dependent $\Gamma_{osc} \sim \delta m^2/E$
the low energy neutrinos start to oscillate first,
and later the
oscillations become noticeable for  the more energetic neutrinos. Due to
that,
the spectrum of the neutrinos (antineutrinos) may become strongly
distorted,  especially in  the case of oscillations
into less abundant  sterile neutrinos. This effect was shown
considerable both in the vacuum oscillations case~\cite{dpk}
and matter oscillations case~\cite{NU96}. Spectrum distortion of
the electron neutrinos due to oscillations leads to
overproduction of helium-4, which may be several times larger
than He-4 overproduction due to an additional neutrino type.

(c) Oscillations can lead to a substantial depletion of the 
number densities ($N_{\nu}$ and $N_{\tilde{\nu}}$) of 
 active neutrinos,  
in case they proceed between active and less abundant sterile 
neutrino states.  
This
slows down the weak rates, $\Gamma_w\sim N_{\nu} E^2_{\nu}$, and again
leads to an earlier $n/p$ - freezing and a corresponding increase of the
helium-4 yield. 

This effect was analytically calculated 
and  found to be important for CN first in  
the vacuum oscillation case~\cite{dpk}. 
For  matter oscillations
with great mass differences it was first estimated in 
ref.~\cite{bd1,bd2}, where it was described in terms of  
an effective chemical potential of neutrinos, generated 
due to active-sterile oscillations. Electron neutrino depletion 
was  numerically calculated, without the 
account of spectrum distortion and asymmetry growth in
ref.~\cite{ekt}. Later  these effects were taken into account   
in ref.~\cite{NU96} for the case of small
mass differences ($\delta m^2 \le 10^{-7}$ eV$^2$).

For oscillations with small mass differences the latter two effects
can lead up to 32$\%$ overproduction of helium-4~\cite{bern}.

(d) Oscillations produce neutrino-antineutrino asymmetry, which on
its turn
influences the evolution of the neutrino and antineutrino ensembles and
the oscillation pattern. In the nonresonant case the neutrino-mixing
produced asymmetry was shown to have a negligible role on CN. 
However, in the resonant oscillation case the asymmetry
effect on CN was shown to be  considerable~\cite{NU96,res}.
For the case of small mass differences it was proven that even very
small asymmetries $L \ll 0.01$ 
considerably influence CN through oscillations, and therefore
asymmetry effect on CN 
should be accounted for during asymmetry's full evolution. 

In general,  dynamically produced asymmetry
at small mixing angles   
suppresses oscillations, which
leads to less overproduction of helium-4 in comparison with the case
of CN with oscillations but without the asymmetry account.
 Hence, the  bounds on the
oscillation parameters
 provided without the asymmetry account
are alleviated at small mixing angles~\cite{res,CAPP}.

Vice versa, the presence of a relic asymmetry in the 
nonresonant oscillation case leads to an alleviation of the 
bounds at large mixings due to suppression of
oscillations by asymmetry. While at small mixings, 
due to spectrum resonance enhancement of oscillations caused by the
asymmetry, 
the bounds on oscillation parameters are strengthened~\cite{NP}.

{\it In case of neutrino oscillations, 
a precise description  of the neutrino evolution at each momentum is
necessary, inorder to account for the essential effects of 
spectrum distortion and asymmetry generation due to oscillations. 
To study the oscillation effects (a)-(d) on the 
light element production in CN, a selfconsistent study of neutrino 
and  nucleons evolution is to be provided.}

\section{Updated constraints on oscillation parameters and the solar
neutrino problem}

Inorder to account for all different effects of oscillations
on CN we have made  a  selfconsistent
 numerical analysis of
the kinetics of the oscillating neutrinos and  the  nucleons at 
freeze-out.

\subsection{The kinetics}

For simplicity we discussed a toy model of oscillations just in the
electron sector $\nu_i=U_{il}~\nu_l$, $l=e,s$:
\begin{eqnarray*}
\nu_1 & = & c\nu_e+s\nu_s,\\
\nu_2 & = & -s\nu_e+c\nu_s, 
\end{eqnarray*}
\noindent where $\nu_s$ denotes the sterile electron antineutrino,
$c=\cos(\vartheta)$,
$s=\sin(\vartheta)$ and $\vartheta$ is the mixing angle in the electron
sector, the mass eigen-states $\nu_1$ and  $\nu_2$ are
Majorana particles with masses correspondingly $m_1$ and $m_2$.

We have solved selfconsistently the set of
the following coupled integro differential equations describing the
evolution of the neutrino density matrix $\rho$ and neutron number
densities $n_n$:

\begin{eqnarray}
&&{\partial \rho(t) \over \partial t} =
H p_\nu~ {\partial \rho(t) \over \partial p_\nu} +
\nonumber\\
&&+ i \left[ {\cal H}_o, \rho(t) \right]
+i \sqrt{2} G_F \left(\pm {\cal L} - {\cal Q}/M_W^2 \right)N_\gamma
\left[ \alpha, \rho(t) \right],
\label{kin}
\end{eqnarray}
\begin{eqnarray}
&&\left(\partial n_n / \partial t \right)
 = H p_n~ \left(\partial n_n / \partial p_n \right) +
\nonumber\\
&& + \int {\rm d}\Omega(e^-,p,\nu) |{\cal A}(e^- p\to\nu n)|^2
\left[n_{e^-} n_p (1-\rho_{LL}) - n_n \rho_{LL} (1-n_{e^-})\right]
\nonumber\\
&& - \int {\rm d}\Omega(e^+,p,\tilde{\nu}) |{\cal A}(e^+n\to
p\tilde{\nu})|^2
\left[n_{e^+} n_n (1-\bar{\rho}_{LL}) - n_p \bar{\rho}_{LL}
(1-n_{e^+})\right].
\end{eqnarray}

\noindent where $\alpha_{ij}=U^*_{ie} U_{je}$,
$p_\nu$ is the momentum of electron neutrino,
 $n$ stands for the number density of the interacting particles,
${\rm d}\Omega(i,j,k)$ is a phase space factor and  ${\cal A}$ is the
amplitude of the corresponding process.
The sign plus in front of ${\cal L}$ corresponds to neutrino,
while minus -- to antineutrino.

The initial condition for the
neutrino ensembles in the interaction basis
is assumed of the form:
$$
{\cal \rho} = n_{\nu}^{eq}
\left( \begin{array}{cc}
1 & 0 \\
0 & 0
\end{array} \right)
$$
where $n_{\nu}^{eq}=\exp(-E_{\nu}/T)/(1+\exp(-E_{\nu}/T))$.

${\cal H}_o$ is the free neutrino Hamiltonian.
The `nonlocal' term ${\cal Q}$ arises as an $W/Z$ propagator effect,
${\cal Q} \sim E_\nu~T$.
${\cal L}$ is proportional to the fermion asymmetry of the plasma
and is essentially expressed through the neutrino asymmetries
${\cal L} \sim 2L_{\nu_e}+L_{\nu_\mu}+L_{\nu_\tau}$,
where
$L_{\mu,\tau} \sim (N_{\mu,\tau}-N_{\bar{\mu},\bar{\tau}})/ N_\gamma$
and $L_{\nu_e} \sim \int {\rm d}^3p (\rho_{LL}-\bar{\rho}_{LL})/N_\gamma$.
Due to the different temperature dependence an interesting 
interplay between these two terms during the cooling of the Universe is
observed. 

  The neutron and proton number
densities, used in the kinetic equations for neutrinos,
are substituted
from the numerical calculations of eq.~(2). On the other hand,
$\rho_{LL}$ and $\bar{\rho}_{LL}$
at each integration step of eq.~(2) are taken  from the
simultaneously  performed integration of the set of equations
(1).

These equations  provide simultaneous account of the different
competing processes,
namely: neutrino oscillations, Hubble expansion and weak interaction
processes.

For  the nonequilibrium
active--sterile oscillations 
this is the only acceptable way of description of the asymmetry and
oscillating neutrinos evolution.  Because, in case of
strongly distorted by oscillations neutrino and antineutrino
spectrum,  the asymmetry cannot be described in a degeneracy terms
anymore, and the 
spectrum may strongly differ from an equilibrium spectrum
with shifted  effective temperature~\cite{NU96}.

 The first equation results into a set of coupled nonlinear
integro-differential equations with time dependent coefficients
for the components
of the density matrix of neutrinos. The number of these equations
can be reduced to 6 equations for each momentum mode
of neutrinos and antineutrinos, due to conservation of the total neutrino
number density
in the discussed model.
The spectrum distortion was described by  1000 bins for the
nonresonant case and by up to   5000 bins for the resonant case.
In case the spectrum was described by $N$ bins,
a system of $6N+1$ coupled integro differential equations
was numerically solved.

The numerical analysis was provided
for the characteristic temperature interval [$2$~MeV, $0.3$ MeV]
~\footnote{At higher temperatures $T>2$ MeV the deviations from
the standard CN  are negligible for the discussed 
model of oscillations with small mass differences.}
and the
full set
of oscillation parameters of the active-sterile oscillation model
\cite{NU96}, namely $\delta m^2\le 10^{-7}$ eV$^2$ and
$\sin^2\vartheta \ge 0.001$.
We calculated precisely the $n/p$ - freezing, which is the
essential  for the
production of helium-4, till temperature $0.3$ MeV, and accounted
adiabatically 
for the  decays of neutrons till the start of
nuclear reactions
below $0.1$ MeV.

\subsection*{Updated $\nu_e\leftrightarrow \nu_s$ constraints}

 Spectrum
distortion, neutrino depletion and neutrino-antineutrino asymmetry
 of the oscillating neutrinos,
as well as the  selfconsistent account of the evolution of
neutrinos and nucleons, is essential for estimating  
oscillations effect on CN.
In the last years the constraints in the electron--sterile sector were
updated,  accounting precisely for these 
effects~\cite{PR}--\cite{res},\cite{now2000}.

The combined  constraints for the nonresonant and the resonant case
of electron--sterile oscillation parameters are shown in Fig.~2  
for different values of relative increase of helium-4,
$\delta Y_p=(Y_{osc}-Y_p)/Y_p$.
 On the left-hand side of Fig.~2 the results for the nonresonant
case are presented. The selfconsistent account of neutrinos and nucleons
evolution and the  precise calculation of neutrino depletion and  
spectrum distortion, allowed to strengthen the previous constraints by 
almost an order of magnitude at large mixings.

\begin{figure}[h]
\begin{center}
\epsfig{file=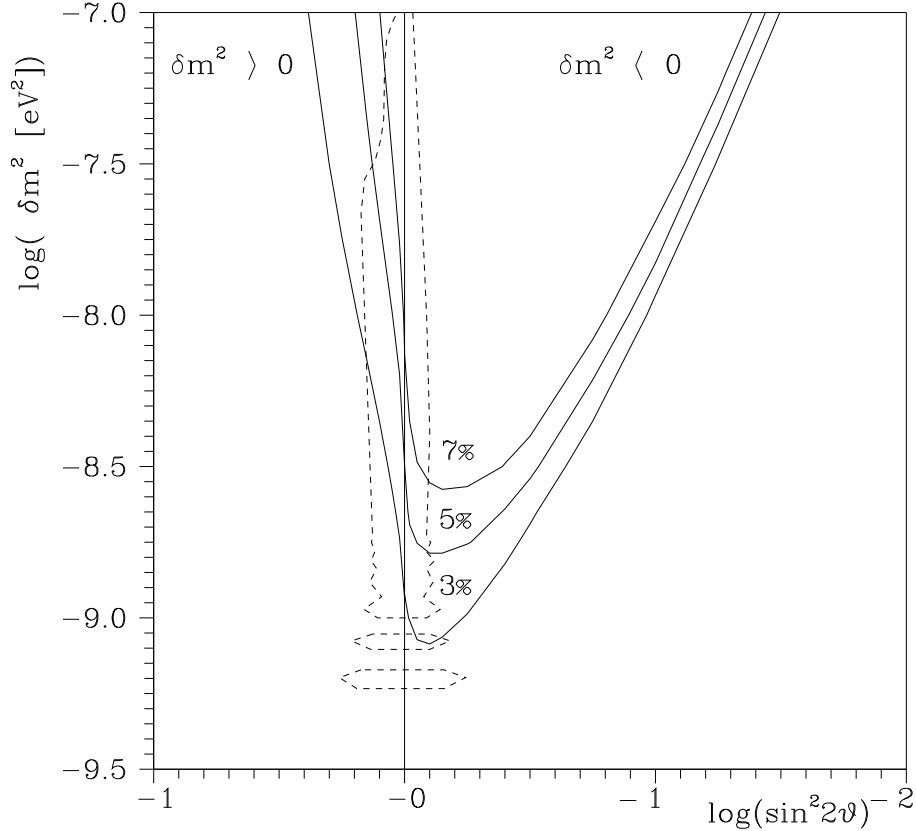,height=11cm,width=12cm}
\end{center}
\vspace{-0.3cm}
\caption{\small The iso-helium contours corresponding
to  $3\%$, $5\%$ and $7\%$
overproduction of primordial helium abundance.
LOW sterile solar solution is given by the
closed dashed curves.}
%\vspace{-0.3cm}   
\end{figure}

The analytical fit to the updated constraints, corresponding 
to  $\delta Y_p=3\%$ is:
\begin{eqnarray*}
&&\delta m^2 (\sin^2 2\vartheta)^4\le 1.5\times 10^{-9}
~{\rm eV}^2~~~ {\rm at}~~~\delta m^2>0,\\
&&\delta m^2 \le 8.2\times 10^{-10}~{\rm eV}^2~~~{\rm at}~~~\delta m^2<0
~\mbox{{\rm at large mixing}}. 
\end{eqnarray*}

In Fig.~3  the plot corresponding to $Y_p=0.24$ is 
compared with previous constraints: 
the pioneer estimates of refs.~\cite{bd1,bd2},
partially accounting for  neutrino depletion; the numerical 
calculations~\cite{ekt}, accounting partially for the neutrino
depletion but neglecting the spectrum distortion and 
the dynamical asymmetry; the
results of a
recent analytical study~\cite{do00}, estimating the  spectrum
distortion effect in the approximation of small mixing angles and
large mass differences.

\begin{figure}[h]
\begin{center}
\epsfig{file=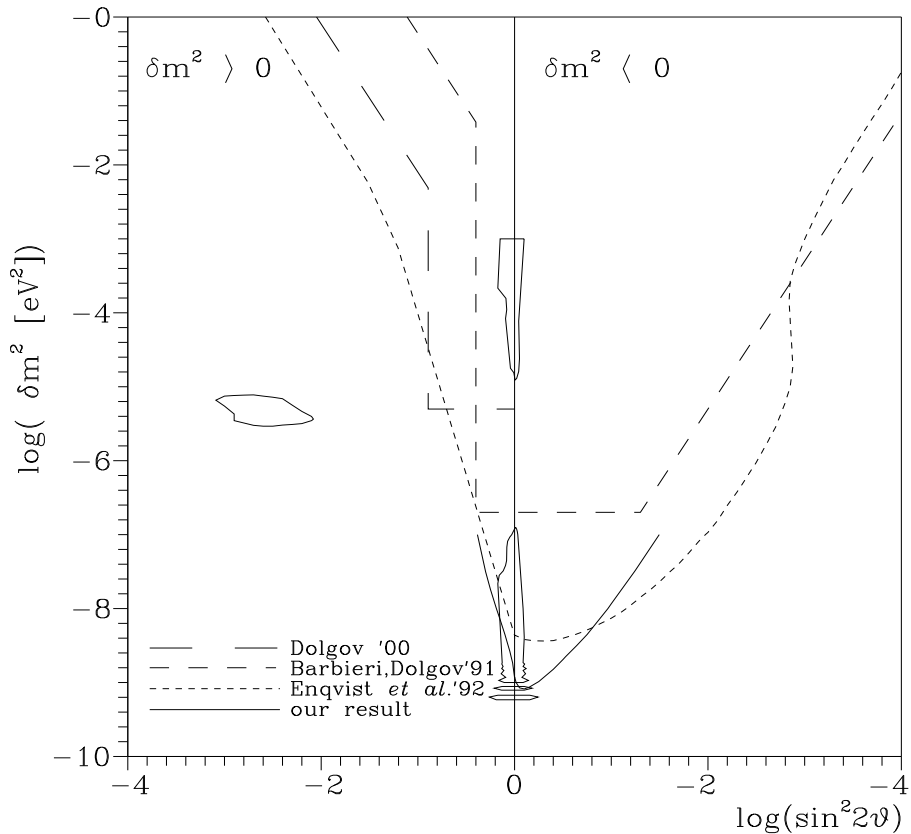,height=11cm,width=12cm}
\end{center}
\vspace{-0.3cm}
\caption{\small Update of the cosmological constraints for the
electron--sterile oscillations. For comparison, the dashed curves show 
the results from  previous
studies neglecting the spectrum distortion and the growth of
asymmetry \cite{bd1,bd2,ekt}, as well as the recent  
analytical constraints\cite{do00} in the nonresonant case.
The precise analysis, accounting for the spectrum distortion and the
asymmetry 
growth~\cite{PR,res} allows an almost complete exclusion of the 
 LOW solution of the solar neutrino puzzle.}
%\vspace{-0.3cm}
\end{figure}

In the resonant case, due to the asymmetry
growth account, He-4 
overproduction is not
so strongly expressed, and hence the old bounds on the oscillation
parameters~\cite{ekt},  provided neglecting  the asymmetry,
were alleviated at small mixing angles (r.h.s. of Fig.~3).
However,  the precise account of the spectrum distortion of      
the oscillating neutrino, and  the exact kinetic approach to both
the neutrino evolution and to the nucleons freeze-out, strengthen
the cosmological constraints at large mixing
 by an order of magnitude.

It will be appropriate to provide similar investigations for  large
mass differences in the electron--sterile case and also to update the
available
constraints for the $\nu_\mu \leftrightarrow \nu_s$ and
$\nu_\tau \leftrightarrow \nu_s$ cases by taking into account the energy
spectrum distortion and the asymmetry growth.
We expect that the constraints on the oscillation parameters for
the $\nu_{\mu,\tau}\leftrightarrow \nu_s$
cases will be more slightly influenced than in the electron--sterile
case,
since $\mu$ and $\tau$ neutrinos do not directly
participate in the
nucleon kinetics.

% but effect only the cooling
%rate of the  Universe and the weak neutrino rates

\subsection*{CN constraints on solar neutrino solutions}

These  constraints on active--sterile
neutrino oscillations exclude  the
active--sterile LOW solution
to the solar neutrino puzzle in addition to the LMA 
solution, excluded in the pioneer works.
 LOW electron--sterile
solution was obtained from the  analysis of
the  1258 days SuperKamiokande experimental data on 
neutrino electron scattering and zenith angle variations 
of the solar neutrino flux~\cite{suzuki,combined}.
It is shown in Figs.~2 and 3 
by the closed solid line
around maximal mixing and $\delta m^2 \sim 10^{-8}$ eV$^2$.

Assumed $\delta Y_p<3\%$,   
electron-sterile LOW solution is almost completely excluded
for $\delta m^2 <0$ and it is 
completely excluded for  $\delta m^2 <0$ case.
It is interesting to note also that 
even in case of very high primordial helium-4  $\delta Y_p/Y_p =
7\%$, sterile LOW 
solution still remains partially excluded~\cite{now2000}. 

According to the global  analysis of the
solar neutrino data from SuperKamiokande, SAGE,
GALLEX + GNO and the Chlorine
experiments, the LMA and  LOW solutions were found not
acceptable for oscillations into sterile 
neutrinos~\cite{garcia,suzuki,krastev}.
The recent global  analyses  
confirmed that LMA and LOW sterile solutions are
disfavoured~\cite{combined,conca}.

The conclusions from the global analysis of 
experimental data in 2000~\cite{garcia,krastev,suzuki} and 
in 2001~\cite{combined,conca} are in 
remarkable agreement with the cosmological constraints 
on  LMA (dating from the early 1980's)
 and on LOW solutions (obtained in 1999).

\ \\

We would like to thank Volodja Karamyshev, Prof. Tsvetan Vylov and the
Organizing
Committee of the
International Workshop on ``Hot points in Astrophysics'', held at JINR,
Dubna, Russia,  for the financial support of the participation of D.K. 
in the Workshop. 

We wish to thank A. Dolgov for useful suggestions and criticism.
We are grateful to   
Rozka Atanassova for the overall help during the preparation of this 
paper.

\end{document}